\documentclass{article}

\usepackage{arxiv}

\usepackage[utf8]{inputenc} 
\usepackage[T1]{fontenc}    
\usepackage[hidelinks]{hyperref}       
\usepackage{url}            
\usepackage{booktabs}       
\usepackage{amsfonts}       
\usepackage{nicefrac}       
\usepackage{microtype}      
\usepackage{graphicx} 
\title{Fast Data-Driven Modeling of Hydraulic Clutch Control Pressure with Latch-State Classification and Gaussian Process Regression\thanks{This work was accepted to the 14th CTI Symposium and Exhibition, ``Automotive Drivetrains, Intelligent, Electrified,'' scheduled for May 13-14, 2020 in Novi, Michigan, USA.}}

\date{May 2020}

\author{
  Yash Bagla \\
  Graduate Controls Engineer\\
	Drive System Design\\
	Farmington Hills, Michigan, 48335\\
    \texttt{yashbagla321@gmail.com} \\
   \And
 Jason Schneider \\
  Senior Controls Engineer\\
	Drive System Design\\
	Farmington Hills, Michigan, 48335\\
  \texttt{Jason.Schneider@drivesystemdesign.com}
}

\begin{document}
\maketitle

\begin{abstract}
This paper presents a data-driven method for modeling the pressure response of a hydraulic clutch control circuit. The system consists of a variable-force solenoid, accumulator, pressure regulator valve, and latch valve, and exhibits nonlinear behavior caused by hysteresis, latch transitions, and actuator dynamics. A baseline model using commanded current variables captured the general pressure response but failed to represent hysteresis and latch behavior accurately. The input vector was therefore extended with current derivative information, and several classifiers were tested to separate latch-related operating regimes before fitting Gaussian Process regression models to the resulting partitions. Nonlinear SVC and gradient boosting produced the highest latch-classification accuracy, and nonlinear SVC was selected for the final local-regression pipeline. The proposed approach was evaluated on unseen ramp-rate data and compared against a physics-based Amesim model. The machine-learning model reproduced the measured pressure response and hysteresis behavior more accurately than the physics-based simulation for the tested operating conditions. These results suggest that machine-learning plant models can complement physics-based hydraulic models during hardware development and controller calibration when representative test-stand data are available.
\end{abstract}


\section{Introduction}

Modern hydraulic control systems are difficult to model accurately because their measured behavior depends on coupled mechanical, hydraulic, and actuator dynamics. In clutch control circuits, effects such as valve friction, hysteresis, latch transitions, dither settings, and pressure-regulator dynamics can cause the same nominal current command to produce different pressure responses under different operating histories. These nonlinearities make calibration and plant-model development expensive when the model must be tuned manually.

Physics-based hydraulic simulation remains valuable during early design because it can be built before hardware is available. However, high-fidelity correlation can require difficult tuning of lumped parameters such as flow friction and jet-force coefficients, and some local nonlinear effects are hard to capture without extensive test data \cite{merritt1967hydraulic,watton1989fluid,jelali2003hydraulic}. Once hardware data are available, a learned plant model can provide a complementary path: it can use measured input-output behavior to reproduce the system response quickly enough for calibration, control development, and sensitivity studies.

This study focuses on a hydraulic clutch control circuit within a valve body for an automatic transmission. The measured output is clutch pressure, and the primary input is the current applied to a variable-force solenoid (VFS), including dither amplitude and dither frequency. The objective is to learn a pressure model that captures the dominant nonlinear behaviors observed in test-stand data, especially hysteresis and latch behavior.

The contribution of this paper is a staged modeling process for this hydraulic subsystem. First, a baseline regression model is trained using commanded current variables. Second, current derivative information is added to represent the direction-dependent hysteresis behavior. Third, multiple classification methods are evaluated for identifying latch-related regimes, and separate Gaussian Process (GP) regression models are trained for the resulting data partitions. The learned model is then compared against measured data and an Amesim physics-based simulation on an unseen ramp-rate dataset. The analysis shows how feature design, regime classification, and local nonlinear regression can be combined to model behaviors that are difficult to capture with a single smooth pressure map.

The remainder of the paper is organized as follows. Section \ref{sec:background} summarizes the relevant learning methods. Section \ref{sec:problem-setup} describes the hydraulic subsystem and test data. Section \ref{sec:algorithm} presents the modeling process. Sections \ref{sec:results} and \ref{sec:conclusions} discuss the results and conclusions.

\section{Background} \label{sec:background}

Nonlinear systems are inherently hard to model because environmental and system parameters can change in ways that are difficult to measure directly. Model predictions for these systems often differ significantly from real-world behavior. Classical system identification and nonlinear black-box modeling provide a framework for learning dynamic system behavior from measured data \cite{ljung1999system,nelles2001nonlinear}, but the model structure and input representation must still be chosen carefully. Complex physics-based models can also be developed and correlated to achieve sufficient accuracy, but this requires information about parameter variability that may be difficult to observe, such as friction forces acting on spool-valve surfaces. Therefore, it is useful to develop techniques that incorporate nonlinear behavior without adding excessive computational or physical measurement cost.

The need to handle uncertainty and nonlinear behavior also appears in autonomous-system domains. For example, chance-constrained receding-horizon planning has been used to reason about uncertain multi-agent motion while balancing collision risk and path feasibility \cite{bagla2019receding, johnson2024impact}. Although the hydraulic-control problem studied here is different, it similarly requires a modeling structure that can represent nonlinear operating behavior without reducing the system to a single nominal trajectory.

The problem of learning a complex model can be addressed by methods such as Locally Weighted Projection Regression (LWPR) \cite{vijayakumar2005incremental}~\cite{schaal2002scalable}. LWPR uses a combination of local linear models to approximate nonlinear functions. This can reduce computation relative to a single global nonlinear model, but it also requires manual tuning of hyperparameters and did not provide the desired accuracy for this application.

Gaussian Process Regression (GPR) provides an alternative regression method in which kernel hyperparameters can be optimized using the marginal likelihood \cite{rasmussen2006cki}. GPR is attractive for this problem because it can represent nonlinear functions and provides a probabilistic prediction framework. Its main limitation is computational cost: exact GP training scales as $\mathcal{O}(n^3)$ with the number of training points $n$. Kernel selection is also important because the kernel encodes assumptions about the function space being learned.

To mitigate this cost and improve local model fidelity, the training data were partitioned into smaller operating regimes before fitting GP regression models. This reduces the matrix size for each local regression problem and allows different operating behavior to be modeled without forcing one global GP to capture all regimes.

This approach follows the motivation of local model learning \cite{nguyen2009local}: instead of using one global model for the full operating range, the data are separated into regions with distinct behavior and separate regression models are trained. In this work, that separation is driven by a classification step designed to identify latch-related behavior in the hydraulic subsystem.

Supervised learning is used here for both classification and regression \cite{bishop2006pattern,hastie2009elements}. Classification separates the measured data into operating regimes, while regression predicts the continuous clutch-pressure response within those regimes. The following sections summarize GPR and Support Vector Classification (SVC), which were used in the final modeling pipeline.

\subsection{Gaussian Process Regression}

When the limited flexibility of a linear model does not provide accurate results, GPR can be a useful algorithm for predicting the behavior of a nonlinear system from measured data. Let a training data set $\mathcal{D}$ contain $n$ observations, $\mathcal{D} = \{(x_i, y_i)\}_{i=1}^{n}$, where $x_i$ denotes the input vector and $y_i$ denotes the measured target value. The objective is to learn a function $f(x_i)$ that predicts $y_i$ such that $y_i = f(x_i) + \epsilon_i$, where $\epsilon_i$ represents Gaussian noise with zero mean and variance $\sigma^2$ \cite{rasmussen2006cki}.

A Gaussian process can define a distribution over functions $f(x)$:
\begin{equation}
    f(x) \sim \mathcal{GP}(m(x), k(x, x'))
\end{equation}

where, m(x) is a mean function given by:

\begin{equation}
    m(x) = \mathbb{E}[f(x)]
\end{equation}

and k is a Rational quadratic or a Radial-basis function (RBF) kernel, used as a covariance function.

Rational Quadratic Kernel
\begin{equation}
    k(x_i, x_j) = \bigg(1 + \frac{d(x_i, x_j)^2}{2 \alpha l^2}\bigg)^{-\alpha}
\end{equation}

 Radial-basis function (RBF) kernel:
 
\begin{equation}
    k(x_i, x_j) = exp\bigg(-\frac{1}{2}{d(x_i/l, x_j/l)^2}\bigg)
\end{equation}

where $l$ is the kernel length scale and $\alpha$ is the scale-mixture parameter \cite{scikit-learn}. In this work, candidate kernels were evaluated empirically and the Rational Quadratic Kernel was selected for the final local GP models.

\subsection{Support Vector Classification}

Support Vector Classification was used to separate operating regions associated with the latch behavior \cite{cortes1995support}. Related support-vector methods are widely used for nonlinear classification and regression when kernel functions are useful for separating nonlinear patterns \cite{smola2004tutorial}. This classification step allowed the regression problem to be divided into smaller and more physically consistent regions before fitting the GP models. In practice, this reduced the burden on a single global regression model and improved the model behavior around the latch transition.

\section{System Setup} \label{sec:problem-setup}

\begin{figure}[!htbp]
\centerline{\includegraphics[width=4.55in]{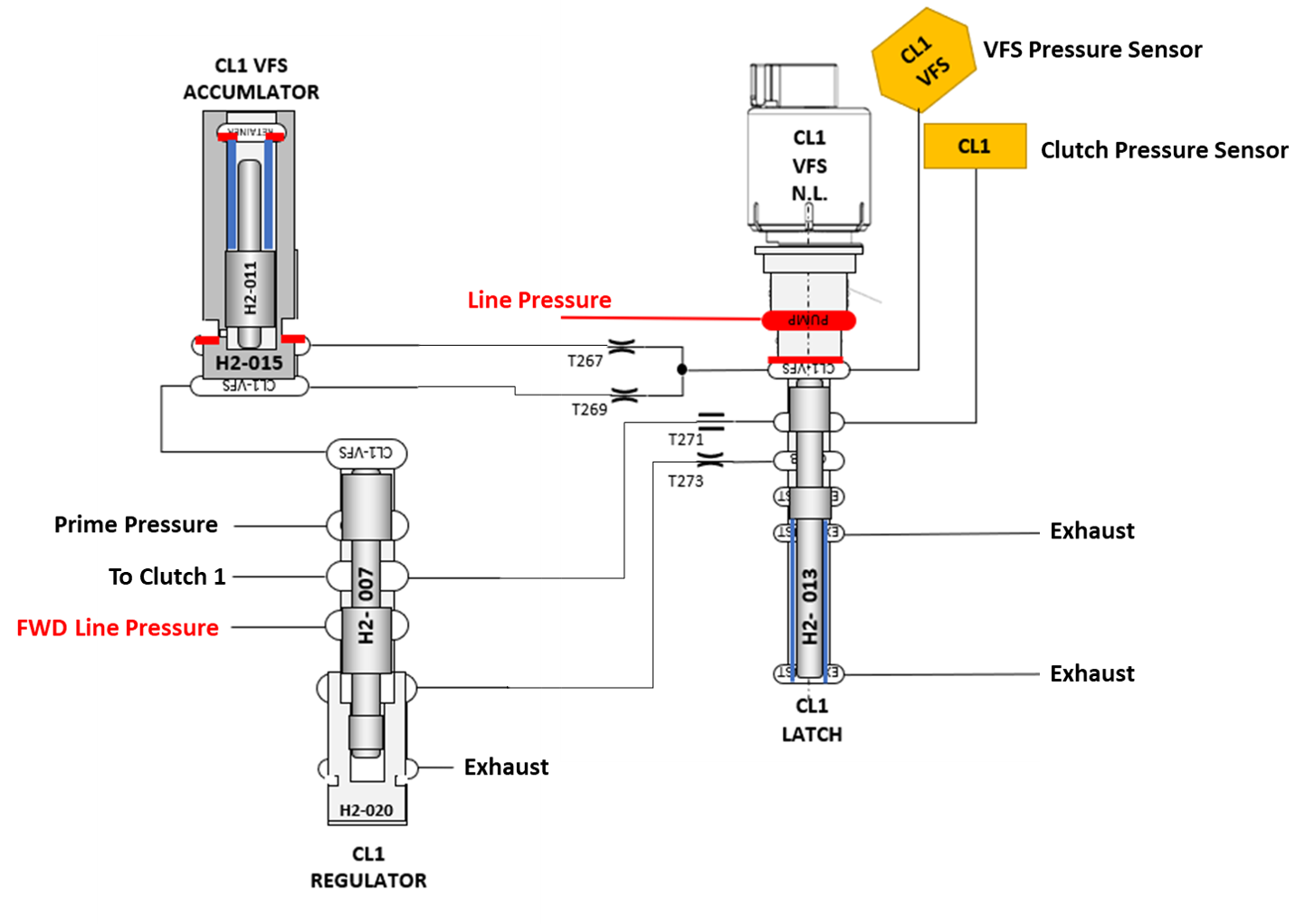}}
\caption{Hydraulic schematic depicting the subsystem of interest. A normally low (NL), variable force solenoid (VFS) is controlled via a PWM current source, ultimately piloting a clutch control valve through the VFS accumulator. The latch valve forces the clutch pressure to be equal to FWD Line Pressure when the VFS pressure rises above a certain threshold by exhausting the feedback port of the clutch regulator valve.}
\label{schematic}
\end{figure}

The test article used to collect training data was a DSD-designed hydraulic valve body for a 5-speed automatic transmission. The data were collected on DSD's hydraulic test stand. This study focuses on a smaller subsystem within the larger valve body: a single clutch control circuit. The subsystem is composed of four main components: a VFS solenoid, an accumulator, a pressure regulator valve, and a latch valve, as shown in Fig. \ref{schematic}.

The primary input variable is the current applied to the VFS solenoid. This current command is defined by the nominal current, dither frequency, and dither amplitude. Dither is commonly applied to current control in these systems to reduce stiction, improve controllability, and reduce hysteresis. The measured output is clutch pressure. In this test setup, the clutch circuit is connected to an accumulator, not pictured in Fig. \ref{schematic}, to replicate the damping that a real clutch would provide. The resulting clutch pressure is measured at a nearby dead-headed port connected to the clutch circuit.

An Amesim model containing representative system geometries, parameters, and fluid properties was created for this system. The simulation includes detailed models of individual spool valves, hydraulic flow paths, and a VFS model obtained directly from the VFS supplier. Default values for lumped parameters such as flow friction and jet-force coefficients were left unchanged from Amesim's default settings.

\section{Learning the Model} \label{sec:algorithm}

\subsection{Data Collected}
The data collected from the valve body test stand consisted of three commanded variables, VFS current, dither amplitude, and dither frequency, and two measured pressure variables, prime pressure and supply pressure. Current was applied to the solenoids at varying ramp rates from 0 mA to 1100 mA. The dither frequency was varied between 30 Hz and 70 Hz, and the dither amplitude was applied as a function of VFS current command with a range of 0 mA to 225 mA. Several variations of the dither amplitude function were tested. The prime and line pressures were largely constant, varying only slightly from values of ($0.5 \pm 0.1$) bar and ($17.7 \pm 0.3$) bar, respectively.

\subsection{Process Development}
The process was developed in several stages as specific deficiencies were identified in the learned model. Initially, the three current-command variables were used as inputs and clutch pressure was used as the output. This baseline model captured the general pressure response, but it did not represent hysteresis well and the latch behavior was poorly captured. This limitation was expected because the input vector did not contain information that indicated direction of travel or latch state.

To address this limitation, the derivative of current was added as an input to indicate the direction of the current ramp. After training with both the absolute current values and the derivative of current, the model captured the observed hysteresis more accurately. However, inaccuracies remained near the latch point.

\begin{figure}[!htbp]
\centerline{\includegraphics[width=5.35in]{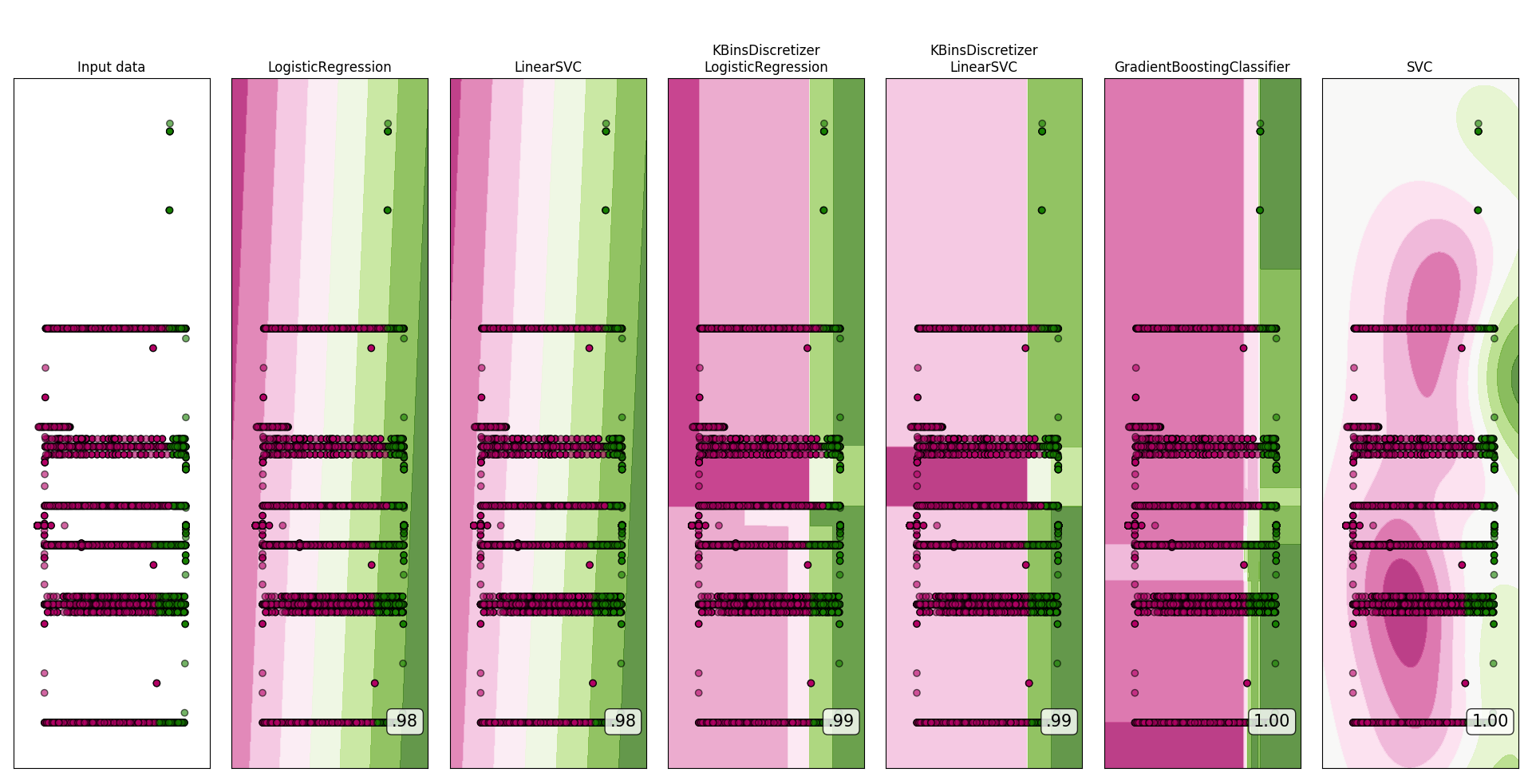}}
\caption{Performance of different classifiers for separating the data set into latched (green) and unlatched (magenta) points. The number in each panel represents the overall classification accuracy.}
\label{class}
\end{figure}

The data were then partitioned so that points before and after latch were treated separately from points associated with the latch event. Several classification techniques were assessed for this task, including logistic regression \cite{hosmer2013applied}, linear SVC, discretized logistic regression, discretized linear SVC, gradient boosting \cite{friedman2001greedy}, and nonlinear SVC. These methods were selected to compare linear decision boundaries, nonlinear kernel boundaries, and ensemble tree-based boundaries. The results are summarized in Table \ref{tab:classifiers} and illustrated in Fig. \ref{class}.

\begin{table}[!htbp]
\centering
\caption{Classifiers evaluated for latch-state separation. Accuracy values correspond to the classifier comparison shown in Fig. \ref{class}.}
\label{tab:classifiers}
\begin{tabular}{lc}
\toprule
Classifier & Classification accuracy \\
\midrule
Logistic regression & 0.98 \\
Linear SVC & 0.98 \\
Discretized logistic regression & 0.99 \\
Discretized linear SVC & 0.99 \\
Gradient boosting classifier & 1.00 \\
Nonlinear SVC & 1.00 \\
\bottomrule
\end{tabular}
\end{table}

\begin{figure}[!htbp]
\centerline{\includegraphics[width=4.45in]{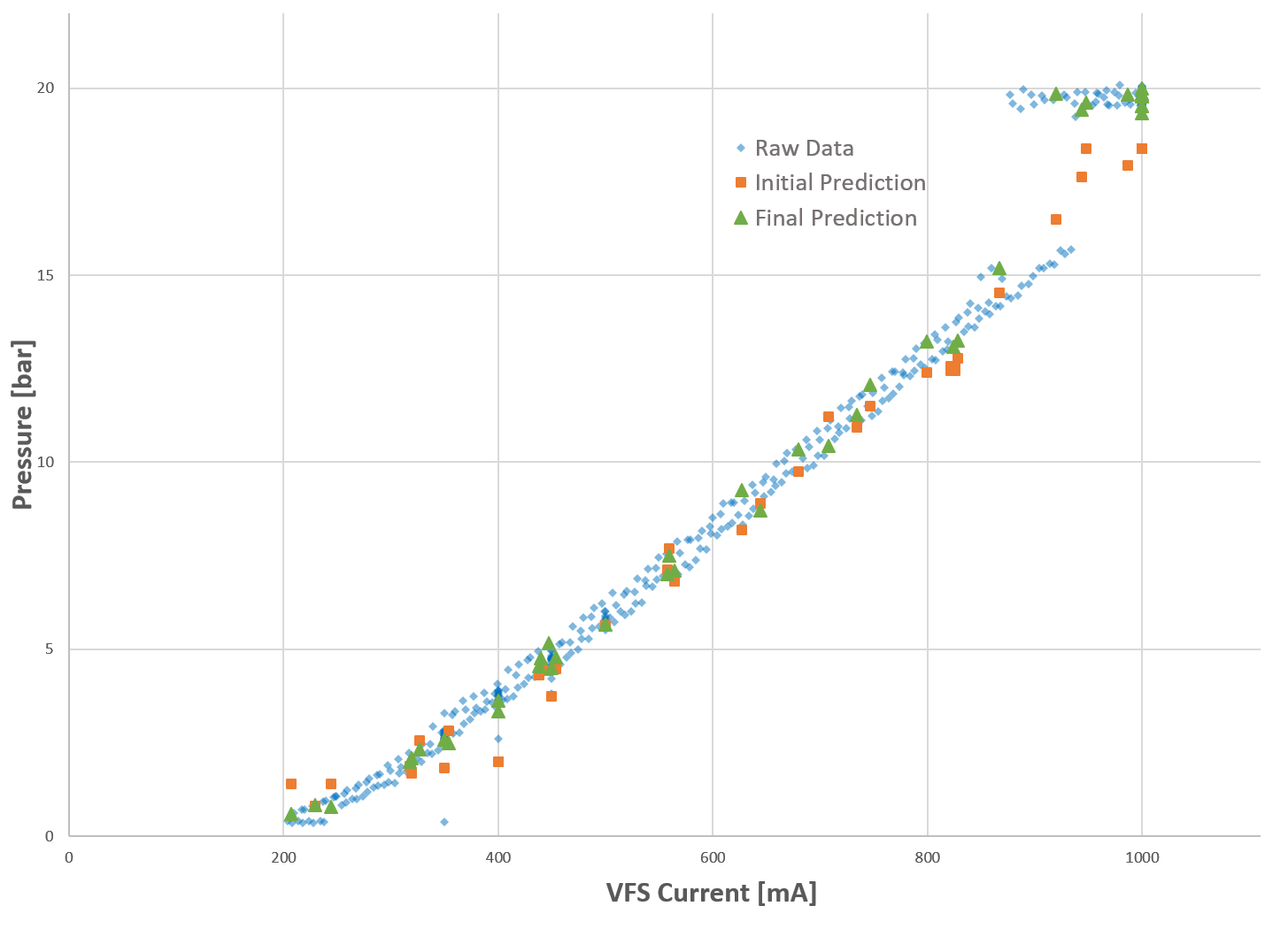}}
\caption{Input-output behavior comparing the original data set, the initial prediction using only VFS current as an input, and the final prediction using current derivative information and SVC-based data partitioning. Blue dots show measured data, orange squares show the initial predictions, and green triangles show the final model predictions.}
\label{predictions}
\end{figure}

Because the nonlinear classifiers performed best on the available data and produced a smooth regime boundary, nonlinear SVC was selected for the classification step. After classification, separate GP models were trained on the resulting data partitions with different candidate kernels, including radial-basis function and Rational Quadratic kernels. The Rational Quadratic Kernel (RQK) produced the best observed behavior in this study. This addition improved prediction near the latch transition. Fig. \ref{predictions} illustrates the evolution of the methodology described above.

The final model should therefore be interpreted as a hybrid supervised-learning pipeline rather than a single algorithm. The classifier identifies the operating regime, the current derivative provides history-direction information for hysteresis, and the local GP models predict the continuous pressure response within the classified regions. This structure was chosen because the latch transition creates a regime change that a single smooth regression model is poorly suited to represent.

\section{Analysis of Results} \label{sec:results}

\begin{figure}[!htbp]
\centerline{\includegraphics[width=4.45in]{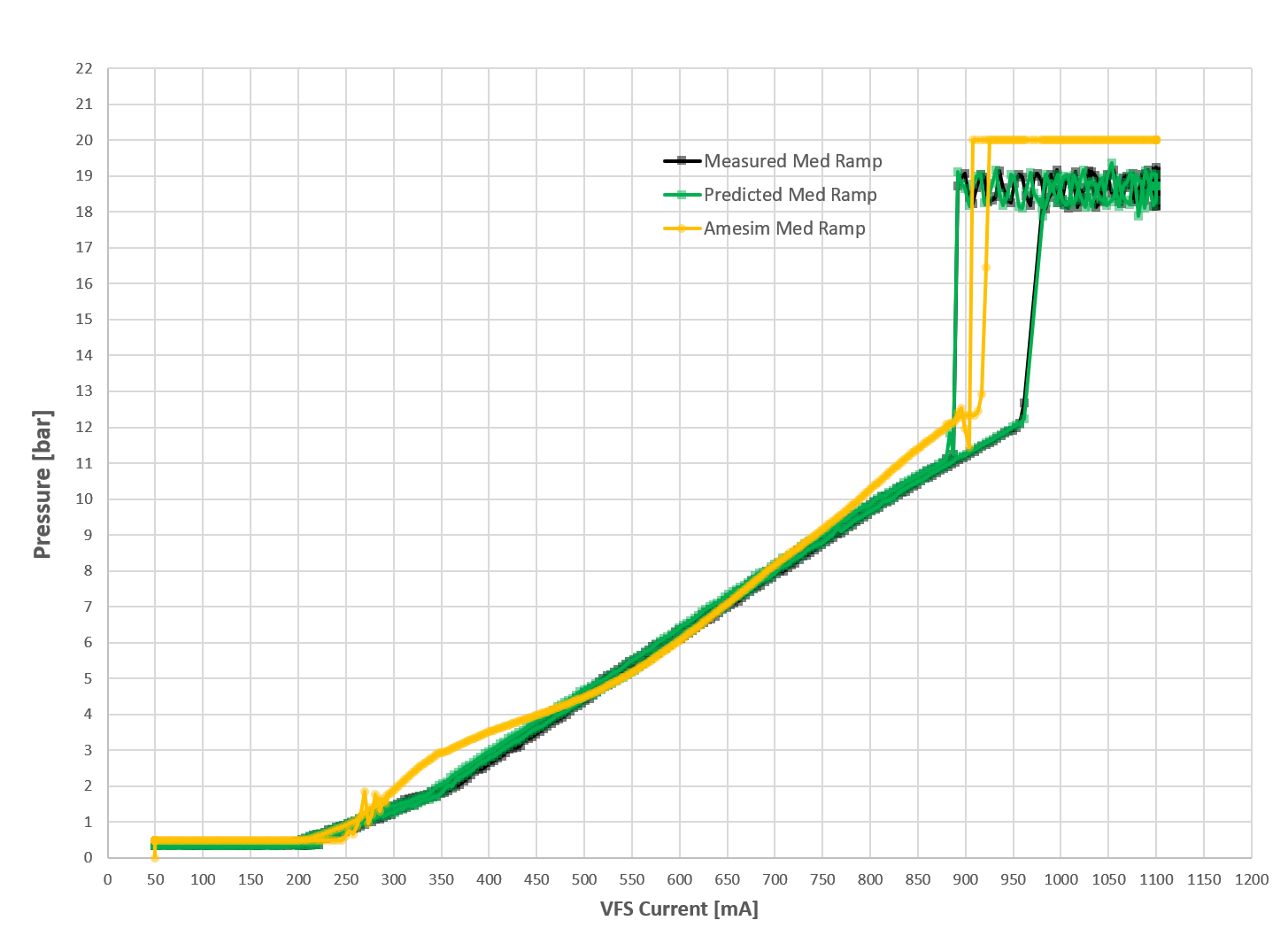}}
\caption{Pressure as a function of VFS current, comparing measured data, Amesim simulation output, and machine-learning predictions for an unseen medium-ramp-rate data set. The black curve shows the measured response, the yellow curve shows the physics-based simulation response, and the green curve shows the response predicted by the machine-learning model.}
\label{MedRamp}
\end{figure}

The final model was trained using data sets where VFS current was ramped up and down at three rates of change, categorized as fast, medium, and slow. Fig. \ref{MedRamp} compares measured data, Amesim simulation output, and machine-learning predictions for an unseen medium-ramp-rate data set.

The evaluation in this paper is based on held-out ramp-rate behavior rather than interpolation on the same traces used for model development. This is important because the primary engineering question is not whether the model can reproduce the training data, but whether the selected input representation and regime partitioning generalize to a measured operating trajectory not used during fitting. The results are presented through pressure-response and hysteresis comparisons because these views directly show the latch transition, rising and falling pressure paths, and ramp-rate-dependent behavior that motivated the model structure.

\begin{figure}[!htbp]
\centerline{\includegraphics[width=4.35in]{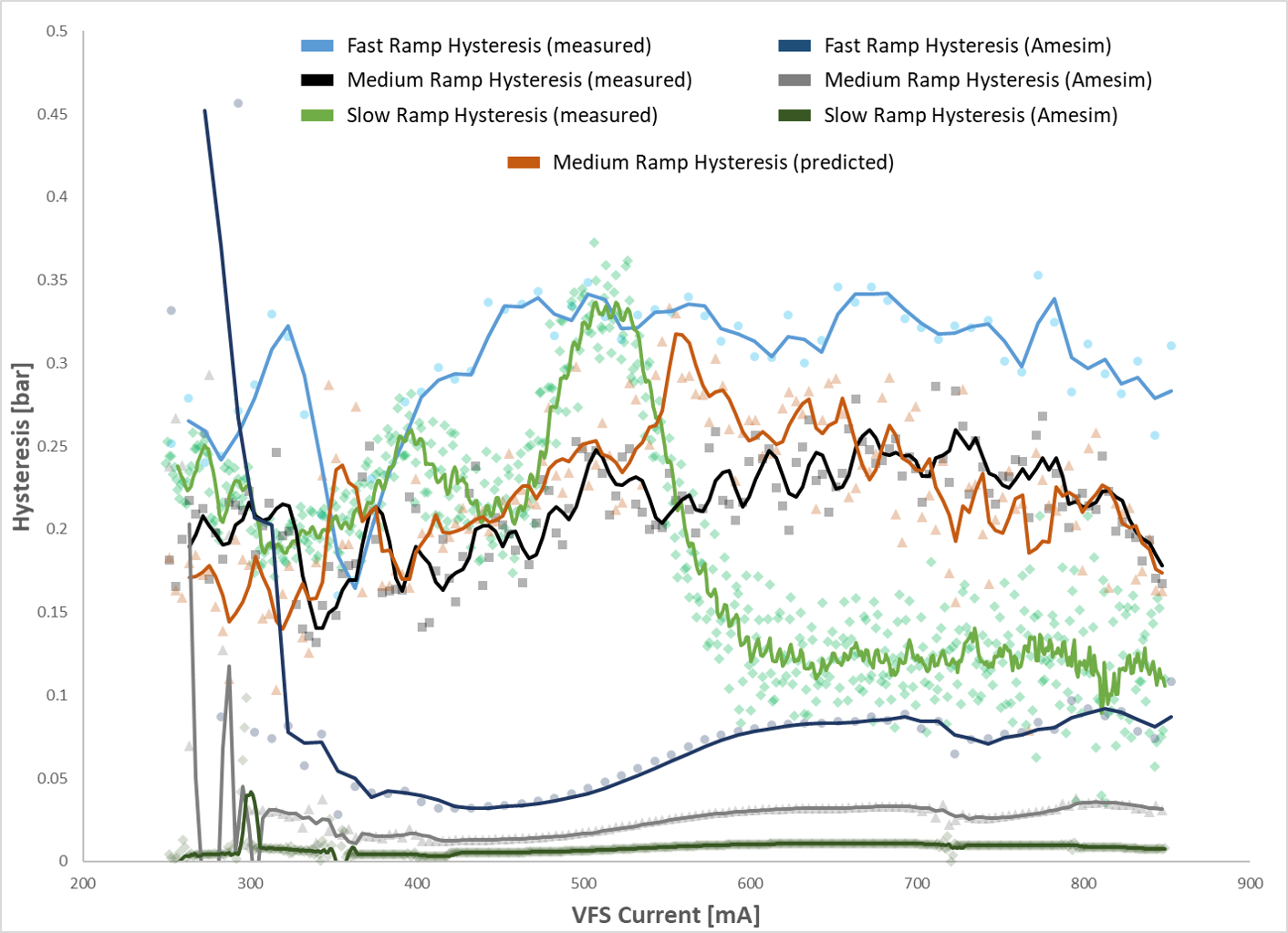}}
\caption{Hysteresis between rising and falling pressure curves as a function of VFS current. The light blue, black, and light green curves show measured fast, medium, and slow ramp-rate hysteresis, respectively. The dark blue, grey, and dark green curves show the corresponding Amesim hysteresis. The brown curve shows the medium-ramp-rate hysteresis predicted by the machine-learning model.}
\label{HystPlot}
\end{figure}

The model reproduced the expected pressure behavior for the evaluated ramp rates. This can be seen most clearly by examining hysteresis between the rising and falling pressure response of the system. Some inherent hysteresis exists because the friction force acting on the outside diameter of the spool valve changes direction when the valve moves to increase or decrease pressure. In addition, system lag between VFS current application, regulator valve motion, and pressure measurement causes different ramp rates to produce different observable hysteresis. The measured hysteresis for all three ramp rates is compared with Amesim simulation output and the machine-learning prediction in Fig. \ref{HystPlot}. The Amesim data were generated by applying the same current commands, including dither content, to the simulated VFS.

The slow-ramp-rate data set exhibits a distinct low-current pressure response before the higher-current hysteresis bands become dominant. Once current exceeds 600 mA, the hysteresis data for the three measured ramp-rate data sets fall into three clearly defined bands that are consistent with the expected ramp-rate-dependent behavior. The Amesim model predicts the trend in hysteresis but underestimates its magnitude and does not correlate closely with the measured data. The machine-learning model more closely follows the measured hysteresis for the unseen medium-ramp-rate data set.

These results show that machine-learning models can be effective for physical hydraulic systems when sufficient representative data can be obtained. The developed input vector accounts for both hysteresis and latch behavior, which were not captured accurately by the baseline model and were not well correlated in the Amesim result shown here. Improved correlation may be obtainable from the simulation software, but this can require extensive tuning of lumped parameters and modeling assumptions. The two approaches therefore support different stages of development. A physics-based model supports early design and validation of operating modes, while a learned model can accelerate calibration and control development once hardware data are available.

The hardware development phase for hydraulic control systems may be reduced by learning from a variety of prescribed test cases. The resulting model can then be used to evaluate the effects of parameters such as orifice sizes, dither settings, or spring constants in a virtual environment before committing to hardware changes. This has the potential to reduce the number of hardware iterations and the time spent making and testing physical modifications.

Another significant benefit of an accurate learned plant model is its potential use in control software development. If the learned model can be implemented with sufficiently low computational cost, it could support feed-forward control development and real-time controller testing.

\section{Conclusions}
\label{sec:conclusions}

This work demonstrates that a machine-learning model can reproduce the pressure response of a hydraulic clutch control circuit when the input representation captures the system behaviors that dominate the measured output. The analysis supports several conclusions. First, commanded-current inputs alone are insufficient to represent the observed hysteresis and latch behavior. Second, adding current derivative information improves the representation of direction-dependent pressure response. Third, separating latch and non-latch operating regions improves the regression problem by allowing separate local GP models to represent different physical regimes. Fourth, nonlinear SVC and gradient boosting produced the strongest latch-classification results among the methods evaluated, with nonlinear SVC selected for the final pipeline because it provided a smooth operating-regime boundary. Compared with the Amesim model used as a physics-based baseline, the learned model better matched the measured pressure response and hysteresis behavior for the unseen medium-ramp-rate case shown in this study.

The results also indicate that physics-based and data-driven plant models are complementary. Physics-based simulation is valuable before hardware exists and can represent design intent, while data-driven models can capture measured nonlinear behavior after test-stand data have been collected. In this development context, the learned model is most useful as a fast plant model for calibration, control development, and sensitivity studies. Future work should evaluate robustness across hardware variation and temperature and test whether the learned model can run in real time on production-relevant controller hardware.

\end{document}